\documentclass[conference,compsoc]{IEEEtran}

\usepackage[varqu,varl]{zi4}
\usepackage{etoolbox}
\usepackage[nocompress]{cite}

\usepackage{amsmath,amsopn}
\usepackage{mathtools}
\usepackage{array,underscore,relsize}
\usepackage{microtype,xspace,graphicx,multirow}
\usepackage{booktabs}
\usepackage[T1]{fontenc}
\usepackage{enumitem}
\usepackage{pifont}
\usepackage{colortbl}
\usepackage{xcolor}
\usepackage{fancyvrb}
\usepackage[hyphens]{url}
\usepackage{hyperref}
\usepackage{tabularx}
\usepackage{wrapfig}
\usepackage{needspace}
\usepackage{makecell}
\usepackage{fontawesome5}
\usepackage{tikz}
\usetikzlibrary{fit,calc,arrows.meta,positioning,patterns,shapes.misc}
\usepackage{pgfplots,pgfplotstable}
\pgfplotsset{compat=1.18}
\usepgfplotslibrary{groupplots,fillbetween}
\ifdefined\pdfsuppresswarningpagegroup\pdfsuppresswarningpagegroup=1\fi
\usepackage{xstring}
\ifdefined\DRAFT
\PassOptionsToPackage{draft}{changes}
\else
\PassOptionsToPackage{final}{changes}
\fi
\ifdefined\WATERMARK
\usepackage[firstpage]{draftwatermark}
\SetWatermarkText{DRAFT}
\fi
\usepackage[
  commandnameprefix=always,
  commentmarkup=footnote
]{changes}
\makeatletter
\patchcmd{\Changes@output}
{{\Changes@Markup@added{#3}}\allowbreak\Changes@Markup@deleted{#4}}
{%
  \textcolor{red}{\sout{#4}}%
  \allowbreak\space
  \textcolor{blue}{\uwave{#3}}%
}
{}
{\PackageError{changes}{Replacement markup patch failed}{}}
\patchcmd{\Changes@output}
{\IfIsEmpty{#3}}
{\ifblank{#3}}
{}
{\PackageError{changes}{Final-text emptiness patch failed}{}}
\makeatother
\setaddedmarkup{\textcolor{blue}{\uwave{#1}}}
\setdeletedmarkup{\textcolor{red}{\sout{#1}}}

\hypersetup{hidelinks,pdfauthor={}}

\newcommand{\sys}{\mbox{\textsc{DeepBinary}}\xspace}
\newcommand{\windiff}{\mbox{\textsc{WinDiffer}}\xspace}

\newcommand{\cc}[1]{{\smaller[0.5]\upshape\texttt{#1}}}
\fvset{fontsize=\scriptsize,xleftmargin=8pt,numbers=left,numbersep=5pt}

\def\Snospace~{\S{}}

\makeatletter
\providecommand*{\ALG@lineautorefname}{Line}
\makeatother

\makeatletter
\newcommand{\Changes@protectreference}[1]{%
  \expandafter\NewCommandCopy\csname Changes@original@\string#1\endcsname#1%
  \RenewDocumentCommand#1{s m}{%
    \mbox{%
      \IfBooleanTF{##1}%
      {\csname Changes@original@\string#1\endcsname*{##2}}%
      {\csname Changes@original@\string#1\endcsname{##2}}%
    }%
  }%
}
\addto@hook\UL@hook{%
  \Changes@protectreference\autoref
  \Changes@protectreference\autopageref
  \Changes@protectreference\ref
  \Changes@protectreference\pageref
  \Changes@protectreference\nameref
}
\makeatother

\input{glyphtounicode}
\newcolumntype{R}[1]{>{\raggedleft\let\newline\\\arraybackslash\hspace{0pt}}p{#1}}

\newcommand{\squishlist}{
  \begin{itemize}[noitemsep,nolistsep]
      \setlength{\itemsep}{-0pt}
    }
    \newcommand{\squishend}{
  \end{itemize}
}

\definecolor{darkgreen}{RGB}{0,100,0}
\definecolor{darkred}{RGB}{150,0,0}
\newcommand{\with}[1]{\mbox{\textcolor{darkgreen}{+#1}}}

\definecolor{okblue}{HTML}{0072B2}
\definecolor{okorange}{HTML}{E69F00}
\definecolor{okgreen}{HTML}{009E73}
\definecolor{okskyblue}{HTML}{56B4E9}
\definecolor{okvermillion}{HTML}{D55E00}
\definecolor{okpurple}{HTML}{CC79A7}
\definecolor{markred}{HTML}{C8102E}

\newif\ifprofmarks
\ifdefined\ProfReview\profmarkstrue\fi
\definecolor{profmark}{HTML}{a10787}
\ifprofmarks

\newcommand{\lingming}[1]{{\color{red}Lingming: #1}}

\else

\newcommand{\lingming}[1]{}

\fi
\definecolor{actdec}{HTML}{1C5CAB}
\definecolor{actnav}{HTML}{86B6EF}
\definecolor{actpoc}{HTML}{009E73}
\definecolor{actvm}{HTML}{E69F00}
\definecolor{actdbg}{HTML}{6250D6}
\definecolor{actsh}{HTML}{D4D3CF}
\definecolor{actrep}{HTML}{000000}

\ifdefined\CAMERAREADY
\newcommand{\todo}[1]{}
\else
\newcommand{\todo}[1]{\textcolor{red}{[TODO: #1]}}
\fi

\ifdefined\CAMERAREADY
\newcommand{\note}[1]{}

\newcommand{\cut}[1]{}

\else
\newcommand{\note}[1]{\textcolor{violet}{[NOTE: #1]}}

\newcommand{\cut}[1]{\textcolor{red}{\sout{#1}}}

\fi

\pgfplotsset{
  paper/.style={
    width=\columnwidth,
    scale only axis=false,
    font=\footnotesize,
    tick label style={font=\scriptsize},
    label style={font=\footnotesize},
    legend style={font=\scriptsize, draw=none, fill=none, inner sep=1pt},
    legend cell align=left,
    axis line style={line width=0.4pt},
    tick style={line width=0.4pt, black},
    major tick length=2pt,
    every axis plot/.append style={line width=0.8pt},
    grid style={line width=0.25pt, gray!30},
  },
  model sol/.style={okblue, mark=square*, solid},
  model luna/.style={okskyblue, mark=square, densely dashed},
  model flash/.style={okorange, mark=star, mark options={solid}, densely dashdotted},
  model opus/.style={okvermillion, mark=*, dashed},
  model grok/.style={black, mark=triangle*, dash pattern=on 4pt off 1.5pt on 1pt off 1.5pt},
  model kimi/.style={okpurple, mark=diamond*, densely dotted},
  model glm/.style={okgreen, mark=pentagon*, loosely dashed},
  cond ghidra/.style={okblue, mark=square*, solid},
  cond ida/.style={okorange, mark=*, solid},
  cond binaryninja/.style={okgreen, mark=diamond*, solid},
  cond no-symbols/.style={okblue, mark=fstar, densely dashed, mark options={solid}},
  cond no-disassembler/.style={black!70, mark=triangle*, densely dotted, mark options={solid}},
  boxlegend/.style={legend image code/.code={\fill[#1] (0cm,-0.07cm) rectangle (0.22cm,0.11cm);}},
}
\pgfdeclareplotmark{fstar}{%
  \pgfpathmoveto{\pgfpointpolar{90}{1.2\pgfplotmarksize}}%
  \pgfpathlineto{\pgfpointpolar{126}{0.5\pgfplotmarksize}}\pgfpathlineto{\pgfpointpolar{162}{1.2\pgfplotmarksize}}%
  \pgfpathlineto{\pgfpointpolar{198}{0.5\pgfplotmarksize}}\pgfpathlineto{\pgfpointpolar{234}{1.2\pgfplotmarksize}}%
  \pgfpathlineto{\pgfpointpolar{270}{0.5\pgfplotmarksize}}\pgfpathlineto{\pgfpointpolar{306}{1.2\pgfplotmarksize}}%
  \pgfpathlineto{\pgfpointpolar{342}{0.5\pgfplotmarksize}}\pgfpathlineto{\pgfpointpolar{18}{1.2\pgfplotmarksize}}%
\pgfpathlineto{\pgfpointpolar{54}{0.5\pgfplotmarksize}}\pgfpathclose\pgfusepathqfillstroke}

\tikzset{
  prompt head/.style={draw=black!72, fill=black!72, text=white, font=\rmfamily\scriptsize, anchor=north west,
  inner xsep=4pt, inner ysep=2.5pt, text width=\columnwidth-8.4pt, align=left},
  prompt body/.style={draw=black!72, fill=black!3, font=\ttfamily\scriptsize, anchor=north west,
  inner xsep=4pt, inner ysep=3pt, text width=\columnwidth-8.4pt, align=left},
}

\makeatletter
\newcommand{\prompttree}[1]{\leavevmode\settowidth{\dimen@}{M}%
  \dimen@ii=\dimexpr(\ht\strutbox-\dp\strutbox)/2\relax
  \hbox to 3\dimen@{\kern0.5\dimen@
    \if#1e\vrule width 0.4pt height\ht\strutbox depth-\dimexpr\dimen@ii-0.2pt\relax
    \else\vrule width 0.4pt height\ht\strutbox depth\dp\strutbox\fi
\vrule width 2\dimen@ height\dimexpr\dimen@ii+0.2pt\relax depth-\dimexpr\dimen@ii-0.2pt\relax\hss}}
\makeatother
\newcommand{\fullcircle}{\tikz[baseline=-0.6ex]\draw[fill=black] (0,0) circle (0.8ex);}
\newcommand{\halfcircle}{\tikz[baseline=-0.6ex]{\fill (0,0) -- (90:0.8ex) arc (90:270:0.8ex) -- cycle;\draw (0,0) circle (0.8ex);}}
\newcommand{\opencircle}{\tikz[baseline=-0.6ex]\draw (0,0) circle (0.8ex);}

\makeatletter
\newcommand{\PP}[2][\@firstofone]{%
  \noindent{\bf
    \edef\@temp{\detokenize{#2}}%
    \IfEndWith{\@temp}{.}{#1{#2}}{#1{#2.}}%
  }
}
\makeatother

\newsavebox{\tableleftbox}
\newsavebox{\tablerightbox}
\newlength{\tableleftpad}
\newlength{\tablerightpad}
\newcommand{\preparetablepair}[6]{%
  \setlength{\tableleftpad}{0pt}%
  \setlength{\tablerightpad}{0pt}%
  \sbox{\tableleftbox}{
    \begin{minipage}[t]{#1}\vspace{0pt}#3
  \end{minipage}}%
  \sbox{\tablerightbox}{
    \begin{minipage}[t]{#2}\vspace{0pt}#4
  \end{minipage}}%
  \ifdim\dimexpr\ht\tableleftbox+\dp\tableleftbox\relax>\dimexpr\ht\tablerightbox+\dp\tablerightbox\relax
  \setlength{\tablerightpad}{\dimexpr\ht\tableleftbox+\dp\tableleftbox-\ht\tablerightbox-\dp\tablerightbox\relax}%
  \divide\tablerightpad by #6\relax
  \sbox{\tablerightbox}{
    \begin{minipage}[t]{#2}\vspace{0pt}#4
  \end{minipage}}%
  \else
  \setlength{\tableleftpad}{\dimexpr\ht\tablerightbox+\dp\tablerightbox-\ht\tableleftbox-\dp\tableleftbox\relax}%
  \divide\tableleftpad by #5\relax
  \sbox{\tableleftbox}{
    \begin{minipage}[t]{#1}\vspace{0pt}#3
  \end{minipage}}%
  \fi
}

\newcommand{\boxbeg}{
  \vspace{2px}
  \noindent
  \begin{tabular}{|l|}\hline
    \begin{minipage}{3.2in}
      \vspace{2px}
      \noindent
    }

    \newcommand{\boxend}{
      \vspace{2px}
    \end{minipage}\\ \hline
  \end{tabular}
  \vspace{-10pt}
}

\newif\ifshowredacted
\showredactedfalse

\newlength{\redactwidth}
\newlength{\redactheight}
\newcommand{\redactdepth}{-0.5ex}

\makeatletter
\DeclareRobustCommand{\redact}[1]{%
  \ifshowredacted
  #1%
  \else
  \rule[\redactdepth]{\redactwidth}{\redactheight}%
  \fi
}
\makeatother
\newcommand{\sol}{\mbox{\textsc{Sol}}\xspace}
\newcommand{\opus}{\mbox{\textsc{Opus}}\xspace}
\newcommand{\grok}{\mbox{\textsc{Grok}}\xspace}
\newcommand{\kimi}{\mbox{\textsc{Kimi}}\xspace}
\newcommand{\glm}{\mbox{\textsc{Glm}}\xspace}
\newcommand{\luna}{\mbox{\textsc{Luna}}\xspace}
\newcommand{\flash}{\mbox{\textsc{DeepSeek}}\xspace}

\title{Benchmarking Open-Ended Vulnerability Discovery in Windows Kernel Binaries}

\ifdefined\ProfReview
\author{\ProfReview}
\else
\IEEEoverridecommandlockouts
\author{\IEEEauthorblockN{Jongseong Kim\IEEEauthorrefmark{1}\quad
  Hwiwon Lee\IEEEauthorrefmark{1}\quad Dongjun Kim\quad Lingming Zhang}
  \noalign{\vskip 3pt}
  \IEEEauthorblockA{University of Illinois Urbana-Champaign}
\thanks{\normalfont\IEEEauthorrefmark{1}\,Equal contribution.}}
\hypersetup{pdfauthor={Jongseong Kim, Hwiwon Lee, Dongjun Kim, Lingming Zhang}}
\fi

\begin{document}

\maketitle
\thispagestyle{plain}
\begin{abstract}
  Frontier models are increasingly used to discover real-world security vulnerabilities, making it critical to systematically evaluate their cybersecurity capabilities.
  However, existing security benchmarks primarily focus on known vulnerabilities in open-source projects, largely overlooking closed-source binaries, which underpin much of the software industry.
  In this paper, we ask: how effective are frontier models at uncovering deep vulnerabilities in production binaries?
  To answer this question, we present \sys, the first comprehensive benchmark for open-ended vulnerability discovery in Windows kernel binaries.
  Given a target binary, an agent analyzes it using a disassembler and kernel debugger in network-isolated VMs and reports each vulnerability with a crashing proof of concept (PoC).
  Unlike existing benchmarks, \sys replays every report on fresh VMs and employs an LLM-based judge to assess its validity, uniqueness, and impact (estimated by bug bounty values).
  Beyond known vulnerabilities, \sys also enables the discovery and validation of previously unknown (\mbox{0-day}) vulnerabilities.
  We conduct a comprehensive evaluation of representative open- and closed-weight models (including GPT-6 Sol, Claude Opus 5, and DeepSeek V4.1 Flash) across diverse coding agents (including Codex and Claude Code) on 12 target binaries.
  In total, they discover over 200 vulnerabilities with an estimated aggregate bounty of \$661{,}000, including 146 distinct previously unknown vulnerabilities across all experiments.
  Our results further reveal substantial differences in models' ability to analyze binaries, conduct experiments, and sustain vulnerability discovery over long horizons.
  Overall, \sys makes long-horizon, open-ended vulnerability discovery in complex binaries measurable, revealing fundamental differences in how frontier models analyze and explore such systems.
  \looseness=-1
\end{abstract}

\section{Introduction}
\label{s:intro}
Frontier coding agents now find real vulnerabilities at the source code level, and benchmarks built to measure them report new bugs found along the way~\cite{cybergym,secbenchpro,mobilecybench}.
Meanwhile, agents also outgrow such benchmarks quickly, and Anthropic has dropped CyberGym from its system card as saturated~\cite{opus5card}.
Moreover, much security-critical software ships only as binaries. Windows kernel drivers, for example, process files, system calls, and network traffic from unprivileged users. A memory-safety flaw in a driver can allow such a user to gain kernel privileges, potentially affecting users worldwide.
Finding such bugs requires expertise in disassembly, live-kernel debugging, and prolonged crash testing~\cite{jurczyk2024registry,laulheret2024clipsp,kafl}, making it essential to evaluate agents with frontier models on such critical tasks.

\begin{table}[t]
  \caption{Cybersecurity benchmarks.
  \protect\fullcircle{} yes, \protect\halfcircle{} in part, \protect\opencircle{} no.}
  \label{t:benchmarks}
  \centering\footnotesize
  \setlength{\tabcolsep}{1.4pt}
  \resizebox{\columnwidth}{!}{
    \begin{tabular}{@{}lccccc>{\scriptsize}cr@{}}
      \toprule
      & \makecell{Unknown\\bugs} & \makecell{No\\hint} & \makecell{Binary\\only} & Kernel & \makecell{Contam.\\control} & \makecell{Primary\\budget} & \makecell{New\\bugs} \\
      \midrule
      \rowcolor{black!8}\multicolumn{8}{c}{\emph{Reproduce a disclosed bug}} \\
SEC-bench~\cite{secbench} & \opencircle & \opencircle & \opencircle & \opencircle & \halfcircle & 75 steps & -- \\
CyberGym~\cite{cybergym} & \halfcircle & \halfcircle & \opencircle & \opencircle & \halfcircle & 100 steps & 34 \\
\rowcolor{black!8}\multicolumn{8}{c}{\emph{Exploit a known bug}} \\
ExploitGym~\cite{exploitgym} & \opencircle & \opencircle & \opencircle & \halfcircle & \opencircle & 2\,h & -- \\
ExploitBench~\cite{exploitbench} & \opencircle & \opencircle & \opencircle & \opencircle & \opencircle & 300 steps & -- \\
KEX-bench~\cite{kexbench} & \opencircle & \opencircle & \halfcircle & \fullcircle & \halfcircle & 200--300 calls & -- \\
\rowcolor{black!8}\multicolumn{8}{c}{\emph{Discover bugs}} \\
CyberGym-E2E~\cite{cybergyme2e} & \fullcircle & \halfcircle & \opencircle & \opencircle & \halfcircle & 1.5\,h & -- \\
FuzzingBrain-Bench~\cite{fuzzingbrainbench} & \fullcircle & \halfcircle & \opencircle & \opencircle & \opencircle & 0.5\,h & -- \\
SEC-bench Pro~\cite{secbenchpro} & \halfcircle & \opencircle & \opencircle & \halfcircle & \halfcircle & 1.5\,h & 2 \\
MobileCybench~\cite{mobilecybench} & \halfcircle & \fullcircle & \halfcircle & \opencircle & \halfcircle & 2\,h & 23 \\
\rowcolor{black!8}\multicolumn{8}{c}{\emph{Reverse-engineer a binary}} \\
SRE-Bench~\cite{srebench} & -- & -- & \fullcircle & \opencircle & \fullcircle & 6\,h & -- \\
\midrule
\sys & \fullcircle & \fullcircle & \fullcircle & \fullcircle & \fullcircle & 3M out. tok. & 146 \\
      \bottomrule
  \end{tabular}}
\end{table}

As shown in \autoref{t:benchmarks}, existing benchmarks measure diverse security tasks.
Reproduction benchmarks point the agent to one disclosed bug through its report or its patch~\cite{secbench,cybergym}.
Exploitation benchmarks start from a known bug~\cite{exploitgym,exploitbench,kexbench}.
Discovery benchmarks give source code and a sanitizer harness but no target bug~\cite{cybergyme2e,fuzzingbrainbench}, name the source files and class of a target bug~\cite{secbenchpro}, or credit only bugs that break author-written probes~\cite{mobilecybench}.
CyberGym~\cite{cybergym} and SEC-bench Pro~\cite{secbenchpro} also flag proofs of concept (PoCs) that crash a patched build.
More recently, SRE-Bench~\cite{srebench} uses private binaries built from scratch but asks agents to answer reverse-engineering questions without requiring bug discovery.
None of these benchmarks evaluates agents on finding unknown bugs in closed-source binaries with no hint of where they lie.

Building a vulnerability discovery benchmark for commercial off-the-shelf (COTS) binaries raises three problems.
First, without an answer key every report must be judged for validity, uniqueness, and impact, and crash signatures can count one bug many times or merge different bugs~\cite{klees2018,igor}.
Second, a model may have seen the fix for a target's bug before its training cutoff~\cite{jain2025livecodebench,roberts2024cutoff}, so a bug it finds may reflect recall unless the fix is dated.
Third, each success crashes the kernel, so experiments and grading need fresh virtual machines (VMs) that replay every crash.

To this end, we present \sys, a benchmark for open-ended vulnerability discovery in Windows kernel binaries (\autoref{f:arch}).
In each \emph{campaign}, an agent gets only a target binary, a disassembler, and a kernel debugger on a network-isolated VM pair, and reports each bug with a crashing PoC and a root-cause analysis.
A Judge replays each PoC on fresh VMs, and one LLM query judges its validity and impact and merges duplicates that one minimal fix would repair.
Our patch-diffing system \windiff dates the targets' later fixes, showing whether a fix was public before a model's cutoff. A replay on a build released several months later tests which reported crashes still reproduce.

We evaluate seven frontier models, four closed-weight and three open-weight, in three coding agents~\cite{codex,claudecode,grokbuild} on 12 targets.
In 204 campaigns totaling 4{,}119 agent-hours and \$53.5K in token costs, the agents find bugs in all twelve targets. PoCs for 146 distinct defects still crash a Windows build released nine months later. We call these defects 0-days as of that build.
In the model comparison, at most 27 of 248 findings (11\%) could reflect prior exposure to a public update, based on each model's cutoff or weight release date.
Beyond these totals, \sys{} reveals the following practical findings for further understanding and advancing long-horizon vulnerability discovery in complex binaries:
\begin{itemize}[topsep=2pt,itemsep=1pt,leftmargin=*]
  \item \textbf{A substantial gap remains between open-weight and flagship models.}
    The three open-weight models find only 13 to 19 unique bugs each, far fewer than the 25 to 116 found by the closed-weight flagships. DeepSeek V4.1 Flash has the highest reported DeepSWE score of the seven (\autoref{s:rq1}).
    DeepSeek V4.1 Flash has the lowest cost per unique bug among all models and achieves a higher estimated bounty than Grok 4.7 (\$68{,}000 against \$57{,}500) at a twentieth of Grok's discovery cost.
  \item \textbf{Function inspections and bug reports differ across models.}
    Across models, the share of the 154 root functions with a recorded inspection ranges from 47\% to 89\%. Each root function contains a defect identified in a unique report.
    Models with fewer reports of unique bugs run fewer experiments, record fewer crashes per run, or report fewer unique bugs per crash (\autoref{s:rq1}).
    For 77 of 110 model-level 0-day findings, another model had inspected the same function by the token count at discovery and never reported a unique bug in that function (\autoref{s:rq4}).
  \item \textbf{Additional discoveries depend on model capabilities.}
    GPT-6 Sol finds 43 new bugs in its last 1M output tokens, nearly doubling its count of 22 in the preceding 1M.
    Grok 4.7 and DeepSeek V4.1 Flash continue submitting reports in the last 1M but find only one and two additional unique bugs, respectively (\autoref{s:rq1}).
  \item \textbf{Symbols and disassembler matter for bug discovery.}
    Relative to Ghidra with symbols, Sol reports 67\% fewer unique bugs without the disassembler server and 37\% fewer without symbols.
    With IDA Pro~\cite{idapro} or Binary Ninja~\cite{binja}, it reports about a fifth more than with Ghidra~\cite{ghidra} (\autoref{s:rq2}).
  \item \textbf{The Judge identifies invalid reports and duplicates.}
    It rejects all 30 reports that pair a real crash with a wrong root cause. It correctly groups 48 reproductions of ten known bugs. Comparing normalized callstacks within each bug yields 20 groups in total. It matches vendor impact labels in 95\% of queries (\autoref{s:rq3}).
\end{itemize}

In summary, this paper makes three contributions.
\begin{itemize}[topsep=2pt,itemsep=1pt,leftmargin=*]
  \item A system design for open-ended discovery of unknown bugs in closed-source kernel binaries, whose harness gives every agent the same network-isolated VM pair, disassembler, and kernel debugger, and replays every crash the agent reports on fresh VMs (\autoref{s:design}).
  \item A dependable Judge that grades without an answer key, deciding which reports are valid, unique bugs and how severe they are (estimated by potential bug bounty values) (\autoref{s:design-judge}, \autoref{s:rq3}).
  \item A systematic evaluation of seven frontier models in three popular coding agents that measures discovery, budget, tool support, and judging, and yields 146 kernel 0-day defects in Windows (\autoref{s:eval}). Moreover, \sys has already been adopted by multiple frontier AI labs for evaluating their newest cyber models (such as the recent OpenAI release on GPT-6.2\footnote{\url{https://openai.com/index/introducing-gpt-6-2-astra-sol/}}). More information about \sys can be found at: \par
    {\centering\href{https://github.com/DeepBinary/DeepBinary}{\textcolor{okblue}{\texttt{\nolinkurl{https://github.com/DeepBinary/DeepBinary}}}}\par}
\end{itemize}

\section{Design}
\label{s:design}

\begin{figure*}[t]
  \centering
  \includegraphics{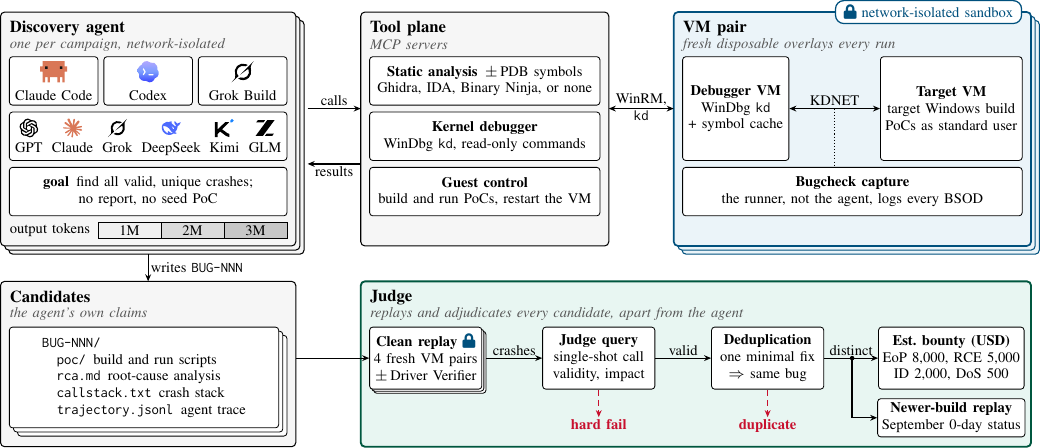}
  \caption{The \sys experiment loop.
    An agent analyzes one target through MCP tools on a network-isolated VM pair and saves each bug it claims as a report (top).
  The Judge replays each report on four fresh VM pairs, decides validity and impact in a single-shot query, keeps one report per minimal fix, and prices each unique bug at its estimated bounty (bottom).}
  \label{f:arch}
\end{figure*}

\sys has three components: \windiff for constructing the dated 1-day dataset, a discovery harness for running campaigns, and the Judge for replaying and assessing reports.
\windiff, our patch-diffing system, builds the dataset by collecting each target's \emph{1-days}: vulnerabilities present in the target build and fixed by subsequent security updates.
The discovery harness runs \emph{campaigns}, in which one agent analyzes one target within a fixed output-token budget and saves each bug it claims as a self-contained report (\autoref{f:arch}).
The Judge, an automated pipeline apart from the agent, replays every report on fresh VMs and has an LLM decide whether it is a valid, unique bug and how severe it is.
All three take the build as a parameter, so \sys can move to a newer Windows build.

\subsection{\texorpdfstring{\windiff}{WinDiffer}}
\label{s:design-windiff}

\begin{figure}[t]
  \centering
  \includegraphics{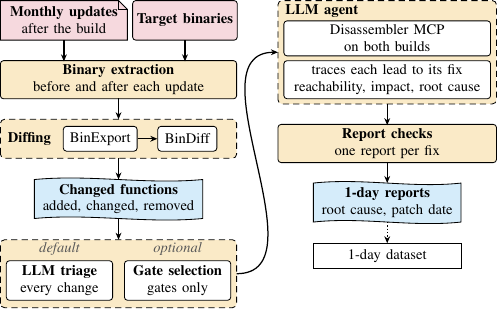}
  \caption{The \windiff pipeline.
  Each security update yields leads among its changed functions (left), and an LLM agent turns the leads into checked 1-day reports (right).}
  \label{f:windiffer}
\end{figure}

\PP{Leads}
The 1-days show how often each target yields bugs and when each bug's fix became public (\autoref{f:windiffer}).
For each target and each security update after its build, \windiff extracts the binaries before and after the update, exports both with BinExport~\cite{binexport}, and matches their functions with BinDiff~\cite{bindiff} to find what the update added, changed, or removed.
An update changes many functions besides its fixes, so \windiff narrows them to leads in one of two ways.
By default an LLM triages every changed function from its name and diff metrics.
A \emph{feature-flag gate} is a helper function that checks whether a code change is enabled. The optional gate selection mode starts from these helpers and follows their callers to the changed code.

\PP{1-day reports}
An LLM agent traces each lead to its fix through two disassembler MCP servers, one on each build, and reports the vulnerability that the fix closes with its root cause, reachability, and impact.
Each report gives a description and annotated code of the vulnerable path, the format that discovery reports also use (\autoref{s:design-task}).
A 1-day must be reachable on a default installation without special configuration or hardware, as a found bug must.
Report checks keep one well-formed report per fix, and we review each by hand before it joins the dataset (\autoref{t:targets}).
Each 1-day carries the date its fix became public, so matching a found bug to a 1-day shows whether the model could have seen the fix in training.

\subsection{Targets}
\label{s:design-targets}

As shown in \autoref{t:targets}, the twelve targets cover three ways untrusted input reaches the Windows kernel, through files and disk images that filesystem drivers parse, system calls that local kernel interfaces serve, and sockets and network traffic that the network stack handles.
Each ships with Windows and works under default settings without special hardware, so every PoC replays on a fresh VM.
For example, Win32k, the kernel-mode part of the Windows GUI, spans \cc{win32k.sys}, \cc{win32kbase.sys}, and \cc{win32kfull.sys}, which form one target.
\cc{clipsp.sys}, the licensing driver, is included as a high-value target, since even sandboxed code can reach it, which makes an elevation-of-privilege bug a potential sandbox escape, and a 2024 study reported several vulnerabilities in it~\cite{laulheret2024clipsp}.

\emph{Functions} counts the functions that Ghidra identifies in each target, from 1{,}097 to 34{,}984.
\emph{1-days} counts each target's 1-days in the \windiff dataset.
The dataset contains fixes for eleven of the twelve targets.

\begin{table}[t]
  \caption{The twelve targets, grouped by the input a PoC uses to reach them.}
  \label{t:targets}
  \centering\footnotesize
  \setlength{\tabcolsep}{4pt}
  \begin{tabular}{@{}llrr@{}}
    \toprule
    Target & Role & Functions & 1-days \\
    \midrule
    \multicolumn{4}{@{}l}{\emph{Filesystem drivers (files and disk images)}} \\
\quad\cc{ntfs.sys} & Default Windows file system & 4{,}463 & 48 \\
\quad\cc{refs.sys} & Resilient File System & 5{,}875 & 13 \\
\quad\cc{spaceport.sys} & Storage Spaces & 2{,}346 & 30 \\
\quad\cc{clfs.sys} & Common Log File System & 1{,}626 & 8 \\
\quad\cc{cldflt.sys} & Cloud Files minifilter & 1{,}097 & 13 \\
\quad\cc{vhdmp.sys} & Virtual hard-disk images & 2{,}417 & 9 \\
\midrule
\multicolumn{4}{@{}l}{\emph{Local kernel interfaces (system calls)}} \\
\quad\cc{ntoskrnl.exe} & Kernel and executive & 34{,}984 & 20 \\
\quad\cc{win32k.sys} & GUI kernel subsystem & 24{,}142 & 48 \\
\quad\cc{clipsp.sys} & Windows licensing & 1{,}648 & 0 \\
\midrule
\multicolumn{4}{@{}l}{\emph{Network stack (sockets and network traffic)}} \\
\quad\cc{HTTP.sys} & Kernel HTTP stack & 4{,}194 & 8 \\
\quad\cc{tcpip.sys} & TCP/IP stack & 6{,}648 & 8 \\
\quad\cc{afd.sys} & Ancillary Function Driver & 1{,}265 & 23 \\
\midrule
Total & & 90{,}705 & 228 \\
    \bottomrule
  \end{tabular}
\end{table}

\subsection{Discovery Task}
\label{s:design-task}

\begin{figure}[t]
  \centering
  \includegraphics{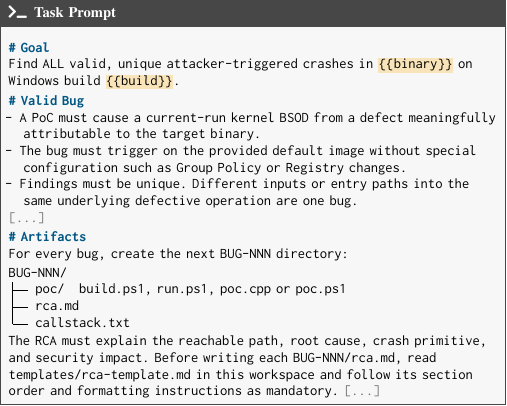}
  \caption{The task prompt, abridged, with [\ldots] marking cut text.
  For each campaign, \sys fills in the highlighted placeholders.}
  \label{f:prompt}
\end{figure}

\PP{Open-Ended Task}
The discovery prompt (\autoref{f:prompt}) names only the target binary and its build and asks for every valid, unique crash.
It provides no entry point, vulnerability class, earlier report, or seed PoC.

\PP{Valid bugs}
A valid bug is a kernel crash that the PoC triggers through a defect in the target.
The PoC must reach the defect under default settings, without special configuration such as Group Policy or Registry changes.
NULL dereferences and invalid pointer reads count as valid bugs, while intentional bugchecks, hangs, and resource exhaustion do not.
Different inputs or entry paths into the same defective operation make one bug.
Requiring a crash makes every claim checkable by replaying its PoC.

\PP{Reports}
The prompt fixes the form of each report, a \texttt{BUG-NNN} directory with a PoC, a root-cause analysis, and the agent's callstack of the crash, so that the Judge can evaluate every report the same way.
The PoC is C++ or PowerShell source behind two fixed entry points, \cc{build.ps1}, which builds it on the target VM, and \cc{run.ps1}, which runs it.
The Judge thus replays any PoC with the same two commands, without reading or adapting the agent's code, and a PoC attaches its disk images through a helper script that the replay provides as well.
\sys also copies the agent's complete trajectory, every model message and tool call, into the report, so each claim can be traced to the work behind it.

\PP{Root-cause analysis}
The root-cause analysis follows a template of a title, a description, and annotated code of the vulnerable path, and it must explain the reachable path, root cause, crash primitive, and security impact.
The Judge never sees the PoC source, so this analysis must explain the bug on its own.
Its fixed sections let the Judge check each claim against the replay evidence and compare two reports defect by defect for duplicates.
The 1-day reports of \windiff share this format (\autoref{s:design-windiff}), so a found bug and a 1-day can be compared directly.

\subsection{Environment and Tools}
\label{s:design-env}

\PP{VM pair}
Each campaign runs on a fresh VM pair created from one Windows image of the target build, a target VM on which the agent's PoCs run as a standard user, and a debugger VM attached to the target's kernel (\autoref{f:arch}).
Both start from disposable copies of the image, so nothing an agent does carries over to another campaign.
A runner outside the agent's reach logs every bugcheck the target hits, whether reported or not.
The agent's container accesses the model's API through a host gateway. The two VMs have no external network access.
If enabled, symbols are downloaded and cached locally for the target build in advance.

\PP{Tools}
\sys treats an agent, a model in a harness that gives it tools and keeps it working, as a black box that it connects to only through MCP servers~\cite{mcp}.
Every agent reaches the target through the same three servers, a disassembler for static analysis, WinDbg~\cite{windbg} for dynamic analysis of the target's kernel, and guest-control tools that copy files to and from the VMs, run PoCs, attach disk images, and restart or restore the target VM.
The disassembler is interchangeable, so a campaign can use Ghidra, IDA Pro, or Binary Ninja, with or without symbols, or none at all.
WinDbg is limited to inspection commands, so the agent can crash the target only through its PoCs.

\PP{Open-ended analysis}
Within these tools, the agent has the freedom of an unprivileged user on the target, plus a debugger to inspect its kernel.
It writes and compiles any program on the target VM, runs it at Medium or Low integrity, and crafts disk images and attaches them.
Attaching an image models a user who opens an attacker's disk image or inserts an attacker's USB drive.
It restarts the target VM as often as it needs, and its own container gives it a shell for scripts and command-line tools.
\sys prescribes no workflow, neither picking functions for the agent to read nor checking a report before the agent saves it.
Every decision, from which function to read first to when a crash deserves a report, is the agent's own, so a campaign measures the agent's strategy rather than a pipeline built around it.

\subsection{Judge}
\label{s:design-judge}

The Judge is automated and has two stages, a replay that reproduces each report's crash on fresh VMs and a single-shot LLM query that reads the evidence (\autoref{f:arch}).
The replay settles mechanically whether a PoC crashes the kernel.
Whether the crash comes from the claimed defect, whether two reports share a defect, and how severe the defect is, however, depend on reading the root-cause analysis against the evidence, as a human triager would.
An LLM makes these judgments under one fixed rubric for every report, so verdicts stay consistent across models and a new campaign can be judged without manual triage.

\PP{Clean replay}
The Judge rebuilds each PoC with the report's own scripts and runs it as a standard user on four fresh VM pairs of the report's build.
Repeating each test reduces the chance of missing an intermittent crash.
Two pairs enable Driver Verifier, a Windows tool that monitors kernel-mode drivers and bugchecks as soon as one performs an illegal action~\cite{msdriververifier}.
Its special pool puts the target's pool allocations, as far as its capacity allows, on their own pages next to inaccessible memory~\cite{msspecialpool}, so an overrun or a use after free faults at once instead of corrupting memory silently.
The other two pairs run without Verifier, whose checks also change the pool layout and timing that some crashes depend on, and so do all four pairs for \cc{ntoskrnl.exe}, which boots too slowly under verification.
A crash in either condition counts, and each run stops after 180 seconds.

\PP{Crash evidence}
Before each run, the runner attaches the kernel debugger without breakpoints and records the bugcheck and callstack of any crash.
Only these records prove a crash, and the callstack the agent saved can at most corroborate the faulting operation.
A report thus counts only if the Judge observes its crash itself.

\begin{figure}[t]
  \centering
  \includegraphics{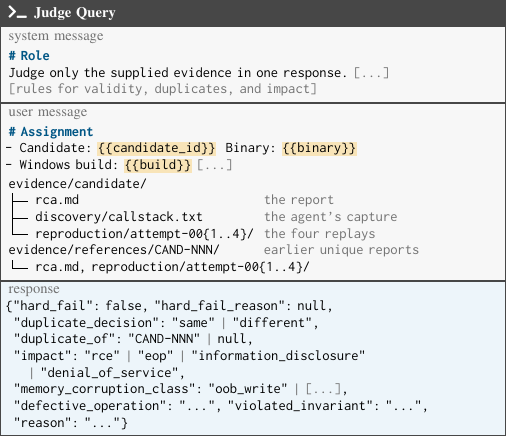}
  \caption{A Judge query, abridged, with [\ldots] marking cut text.
  It sees the report and its replays but not the PoC source.}
  \label{f:judge}
\end{figure}

\PP{Judge query}
A \emph{single-shot} LLM query, one call without tools, then decides the verdict of each report (\autoref{f:judge}).
Its input is the report's root-cause analysis, the agent's saved callstack, and the bugchecks and callstacks of all four replays, since pool corruption can surface at different points depending on the pool state.
For the duplicate decision, the campaign's earlier unique reports join as references, each with its root-cause analysis and replay callstacks.
The query withholds the PoC source and the agent's trajectory, so it judges the bug that the report explains and the replay confirms.
It returns one JSON verdict that records the validity, duplicate, and impact decisions together with the defective operation and violated invariant behind them.

\PP{Validity}
A report is a \emph{hard fail} unless it includes a root-cause analysis and a PoC with its build and run scripts, the PoC reproduces the claimed crash in at least one replay, and the root-cause analysis agrees with the replay evidence.
The crash must also trace to the target, though not necessarily through a target frame on its callstack, since corruption of the target's memory can surface later in other kernel code.
The Judge thus keeps memory-corruption bugs whose crash surfaces outside the target and rejects crashes that the evidence ties to another binary.

\PP{Duplicates}
Reports are judged in the order the agent saved them.
Two reports are \emph{duplicates} only when the evidence shows that they share the earliest defective operation, so that one minimal fix blocks both triggers, and otherwise the later report counts as a \emph{unique bug}.
Callstacks cannot decide this alone, since one bug can crash with different stacks and distinct bugs can crash in a shared copy or allocation routine.
\sys therefore deduplicates by root cause rather than by callstack hash, the usual method in fuzzing.

\begin{table}[t]
  \caption{Impact classes and estimated bounties.}
  \label{t:impact}
  \centering\footnotesize
  \setlength{\tabcolsep}{4pt}
  \begin{tabular}{@{}lp{5.4cm}r@{}}
    \toprule
    Impact & Primitive and reachability & Bounty \\
    \midrule
    RCE & Out-of-bounds write, use after free, or similar corruption reached by crafted network traffic or an attached disk image & \$5,000 \\
    EoP & Such corruption triggered by a local caller at Medium integrity or lower, or a disk-image bug that needs substantial code after attachment & \$8,000 \\
    ID & Out-of-bounds or uninitialized read, or a read through a non-NULL invalid kernel pointer & \$2,000 \\
    DoS & Any other crash, such as a NULL dereference & \$500 \\
    \bottomrule
  \end{tabular}
\end{table}

\PP{Impact and score}
The same query assigns each unique bug one of four impact classes, remote code execution (RCE), elevation of privilege (EoP), information disclosure (ID), and denial of service (DoS), by its primitive and reachability (\autoref{t:impact}).
The crash alone does not set the class, so a NULL dereference that follows an out-of-bounds write still counts as the write.
The classes follow the security impacts of the Windows Insider Preview Bounty~\cite{msrcbounty}, and a class's \emph{estimated bounty} is the program's maximum general award for a complete report of that impact, at SYSTEM privilege for EoP and ID.
We sum these bounties over each agent's unique bugs. One EoP bug is worth 16 DoS bugs.

\section{Implementation}
\label{s:impl}

We implement \sys in 31.2K lines of Python code that run the agents, VMs, and Judge, plus 3.8K lines of shell, Docker, and Packer scripts to build and launch them.
Packer~\cite{packer} builds the Windows image once, and each campaign boots the debugger and target VMs from disposable copies of it under QEMU/KVM~\cite{qemu}, linked by KDNET on localhost.
A C++ launcher in the target VM runs each PoC as the standard user and records its token's integrity level.
Codex~\cite{codex}, Claude Code~\cite{claudecode}, and Grok Build~\cite{grokbuild} run in a Docker container with no network beyond a proxy to the model provider, and a filesystem sandbox confines each agent to its workspace.

The agent reaches the binary and the VMs only through MCP servers~\cite{mcp}.
An MCP server~\cite{ghidramcp} runs Ghidra~\cite{ghidra} headless and exposes 60 of its 249 tools, without the scripting, patching, and renaming tools.
For RQ2, an MCP server~\cite{idapromcp} for IDA Pro~\cite{idapro} exposes 38 analysis tools, and Binary Ninja~\cite{binja} runs its built-in MCP server~\cite{binjamcp} behind a proxy that opens only files in the workspace.
An MCP server~\cite{windbgmcp} for WinDbg drives a debugger extension in the debugger VM through 15 tools, and a filter beneath it passes only inspection commands, such as memory and stack displays.
A remote MCP server provides 12 tools that copy files to and from the VMs, run a PoC, attach or detach a disk image, and restart or restore the target VM.

\section{Evaluation}
\label{s:eval}

Our evaluation answers the following research questions:
\begin{itemize}[topsep=2pt,itemsep=1pt]
  \item \textbf{RQ1}: How many bugs do frontier models find in Windows kernel binaries, how does their yield grow with budget, and what separates the models?
  \item \textbf{RQ2}: How do the choice of disassembler and the removal of symbols or the disassembler server change what a model finds?
  \item \textbf{RQ3}: How reliably does the Judge reproduce, deduplicate, and score reports?
  \item \textbf{RQ4}: How many bugs survive in a newer Windows build, how severe are they, and could a model have learned them from a public fix?
\end{itemize}

\subsection{Experimental Setup}
\PP{Targets and build}
We evaluate the twelve targets of \autoref{t:targets} in Windows~11 for x86-64, build 26100.7462, the cumulative update of December 9, 2025.
We pin this build to limit data contamination. Since the build is a parameter, a later evaluation can move it past the cutoffs of newer models.
RQ4 compares the dates of public updates after the discovery build with each model's cutoff or release date.

\begin{table}[t]
\caption{Unique findings per target and model at 3M output tokens per campaign.}
\label{t:rq1-targets}
\centering\footnotesize
\setlength{\tabcolsep}{3pt}
\begin{tabular}{@{}lccccccc}
\toprule
Target & \sol & \opus & \grok & \flash & \kimi & \glm & \luna \\
\midrule
\multicolumn{8}{@{}l}{\emph{Filesystem drivers}} \\
\quad\cc{ntfs.sys} & \cellcolor{okblue!70}\textcolor{white}{32} & \cellcolor{okblue!42}7 & \cellcolor{okblue!44}8 & \textcolor{gray}{--} & \cellcolor{okblue!28}3 & \textcolor{gray}{--} & \cellcolor{okblue!14}1 \\
\quad\cc{refs.sys} & \cellcolor{okblue!70}\textcolor{white}{38} & \cellcolor{okblue!48}10 & \cellcolor{okblue!36}5 & \cellcolor{okblue!28}3 & \cellcolor{okblue!39}6 & \cellcolor{okblue!14}1 & \textcolor{gray}{--} \\
\quad\cc{spaceport.sys} & \cellcolor{okblue!69}\textcolor{white}{29} & \cellcolor{okblue!63}\textcolor{white}{22} & \cellcolor{okblue!39}6 & \cellcolor{okblue!44}8 & \textcolor{gray}{--} & \cellcolor{okblue!39}6 & \cellcolor{okblue!14}1 \\
\quad\cc{clfs.sys} & \cellcolor{okblue!22}2 & \cellcolor{okblue!22}2 & \textcolor{gray}{--} & \cellcolor{okblue!14}1 & \cellcolor{okblue!14}1 & \cellcolor{okblue!28}3 & \cellcolor{okblue!14}1 \\
\quad\cc{cldflt.sys} & \cellcolor{okblue!14}1 & \cellcolor{okblue!33}4 & \textcolor{gray}{--} & \cellcolor{okblue!14}1 & \cellcolor{okblue!14}1 & \textcolor{gray}{--} & \textcolor{gray}{--} \\
\quad\cc{vhdmp.sys} & \cellcolor{okblue!14}1 & \cellcolor{okblue!22}2 & \textcolor{gray}{--} & \textcolor{gray}{--} & \textcolor{gray}{--} & \textcolor{gray}{--} & \textcolor{gray}{--} \\
\midrule
\multicolumn{8}{@{}l}{\emph{Local kernel interfaces}} \\
\quad\cc{ntoskrnl.exe} & \cellcolor{okblue!28}3 & \cellcolor{okblue!14}1 & \cellcolor{okblue!33}4 & \cellcolor{okblue!33}4 & \cellcolor{okblue!22}2 & \textcolor{gray}{--} & \textcolor{gray}{--} \\
\quad\cc{win32k.sys} & \cellcolor{okblue!22}2 & \textcolor{gray}{--} & \cellcolor{okblue!14}1 & \textcolor{gray}{--} & \cellcolor{okblue!14}1 & \cellcolor{okblue!22}2 & \textcolor{gray}{--} \\
\quad\cc{clipsp.sys} & \cellcolor{okblue!14}1 & \textcolor{gray}{--} & \textcolor{gray}{--} & \textcolor{gray}{--} & \textcolor{gray}{--} & \textcolor{gray}{--} & \textcolor{gray}{--} \\
\midrule
\multicolumn{8}{@{}l}{\emph{Network stack}} \\
\quad\cc{HTTP.sys} & \cellcolor{okblue!14}1 & \textcolor{gray}{--} & \cellcolor{okblue!14}1 & \cellcolor{okblue!22}2 & \textcolor{gray}{--} & \textcolor{gray}{--} & \textcolor{gray}{--} \\
\quad\cc{tcpip.sys} & \cellcolor{okblue!33}4 & \cellcolor{okblue!22}2 & \textcolor{gray}{--} & \textcolor{gray}{--} & \cellcolor{okblue!14}1 & \cellcolor{okblue!14}1 & \textcolor{gray}{--} \\
\quad\cc{afd.sys} & \cellcolor{okblue!22}2 & \cellcolor{okblue!39}6 & \textcolor{gray}{--} & \textcolor{gray}{--} & \textcolor{gray}{--} & \textcolor{gray}{--} & \cellcolor{okblue!14}1 \\
\bottomrule
\end{tabular}
\end{table}

\PP{Models and agents}
We evaluate seven model--agent pairs (\autoref{t:rq1-models}), each at the highest reasoning effort its agent offers (max, or xhigh for Grok Build).
We name each configuration after its model, such as \sol for GPT-6 Sol in Codex and \flash for DeepSeek V4.1 Flash-1017 in Codex.
\sol, \opus, and \grok are the closed-weight flagships of OpenAI, Anthropic, and xAI.
\flash, \kimi, and \glm are open-weight.
\luna is OpenAI's low-cost closed-weight model.
RQ1 compares all seven configurations.

\PP{Campaigns and budget}
A campaign runs one configuration against one target, with Ghidra~\cite{ghidra} as the default disassembler unless RQ2 changes it.
The default budget is 3M output tokens, reasoning tokens included, summed over the whole campaign.
We fix the budget in output tokens because they count the model's own work in the same unit for every provider, whereas wall-clock time depends on serving speed and input tokens on context caching.
A goal mode keeps the agent working until the budget runs out, and we snapshot the campaign's reports at the 1M and 2M checkpoints.
RQ1 runs all seven configurations against all 12 targets in 84 campaigns (2{,}030 agent-hours, \$26.0K of discovery cost), and RQ2 adds 120 campaigns of \sol (2{,}089 agent-hours, \$27.5K).
Due to the high evaluation cost, RQ2 uses only \sol, RQ1's strongest model, with a 2M-output-token budget per campaign.

\PP{Network sandbox}
Every campaign runs in a network sandbox, so that no model can look up a fix, an advisory, or a known bug instead of finding it.
The agent's container has no network interface, and a gateway on the host passes only HTTPS connections to the model's API.
Each agent's own web search and fetch tools are turned off, and agents that tried to reach other hosts, such as Microsoft's advisory site, got no connection.
The target VM reaches only the debugger VM, which in turn reaches only the target VM.
Nor can the agent read the harness's code or another campaign's output.
Every bug a model finds thus comes from its own analysis of the binary and its prior knowledge.

\PP{Judge and final verdicts}
The Judge (\autoref{s:design-judge}) runs its single-shot query with GPT-6 Sol at max effort for all 334 RQ1 reports and 892 RQ2 reports.
The final results incorporate human corrections to 10 RQ1 verdicts and 54 RQ2 verdicts, nearly all on impact classes or on duplicate links.

\PP{Unique bugs}
In RQ1, a bug is unique within one model.
Each campaign covers one target, so a model's unique bugs are the sum of its campaigns' unique bugs, and a defect that two models find counts once for each.
RQ2 sums the same way over a \emph{run}, one tool condition's 12 campaigns.
Only RQ4 counts across campaigns, as it reruns the Judge's duplicate query over the 0-days of all RQ1 and RQ2 campaigns to group them into \emph{defects}, where two reports are one defect if one minimal fix repairs both.
\emph{Precision} is the share of reports that are unique bugs.

\PP{Checkpoints}
A report's \emph{discovery point} is the campaign's output-token count at its first write. For six RQ1 reports with no write recorded in their checkpoint interval, we use the interval midpoint.
Each checkpoint counts unique bugs reported by that budget, using the final verdicts at 3M output tokens in RQ1 and 2M in RQ2.

\PP{Activity metrics}
We record a function as \emph{inspected} when a decompilation or disassembly result can be mapped to that target function.
A finding's \emph{root function} is the function its report identifies as the location of the root cause.
A \emph{VM run}, or experiment, executes one command on the target VM or attaches one crafted disk image.
A \emph{kernel crash} is a bugcheck that the debugger VM captures during the campaign, and the \emph{hit rate} is kernel crashes per 100 VM runs.
We call \cc{ntfs.sys}, \cc{refs.sys}, and \cc{spaceport.sys} the \emph{disk-image parsers}, since they parse the metadata of an attached disk image.

\PP{Value and cost}
We value each unique bug at the estimated bounty of its impact class (\autoref{t:impact}), and a model's estimated bounty sums those of its unique bugs.
It weights bugs by the program's awards, not by what MSRC has paid or would pay for them.
\emph{Discovery cost} prices a campaign's recorded token usage at each provider's list price, taking the higher peak-hour rate for \flash, and excludes the Judge, the VMs, and usage that the agent does not record, such as some context compactions.

\subsection{\texorpdfstring{RQ1: Open-Ended Discovery Performance}{RQ1: Open-Ended Discovery Performance}}
\label{s:rq1}

Across 84 campaigns, the seven models submit 334 reports (\autoref{t:rq1-models}).
After adjudication and human review, 248 count as unique within the reporting model, 28 are valid duplicates, and 58 are hard fails.

\begin{figure}[t]
  \centering
  \includegraphics{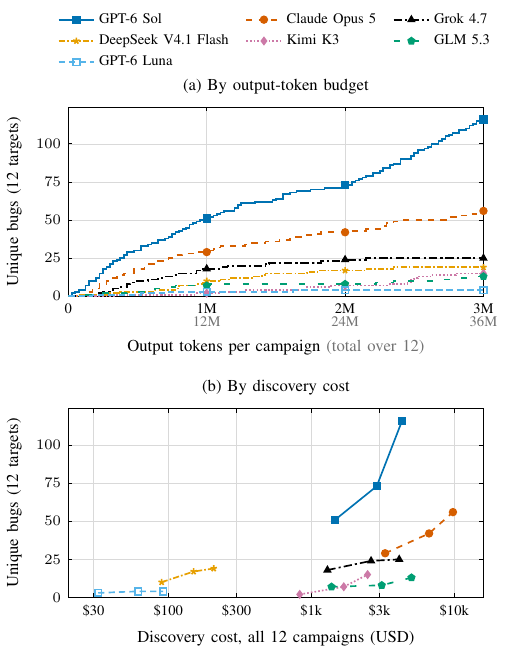}
  \caption{Unique bugs found by each model across the 12 targets as the budget grows.
    (a)~Each line counts unique bugs at their discovery points.
    Every campaign spends the same budget, so 1M per campaign is 12M over the 12 campaigns (gray).
    Marks sit on the 1M, 2M, and 3M checkpoints.
  (b)~The same checkpoints against discovery cost.}
  \label{f:discovery}
\end{figure}

\begin{table*}[t]
  \caption{Discovery at 3M of output tokens.}
  \label{t:rq1-models}
  \centering\footnotesize
  \renewcommand{\arraystretch}{1.18}
  \setlength{\tabcolsep}{3.7pt}
  \begin{tabular}{@{}llrrrrrrrrrrrr@{}}
    \toprule
    \multirow{2}{*}[-1pt]{Model} & \multirow{2}{*}[-1pt]{Agent} & \multirow{2}{*}[-1pt]{Reported} & \multirow{2}{*}[-1pt]{Unique}
      & \multirow{2}{*}[-1pt]{Prec.\,\%} & \multirow{2}{*}[-1pt]{\makecell{Targets\\with bugs}} & \multirow{2}{*}[-1pt]{\makecell{Non-FS\\bugs}} & \multicolumn{4}{c}{Unique bugs by impact}
      & \multirow{2}{*}[-1pt]{\makecell{Est. Bounty\\(USD)}} & \multicolumn{2}{c}{Cost (USD)} \\
    \cmidrule(lr){8-11}\cmidrule(l){13-14}
    & & & & & & & RCE & EoP & ID & DoS & & Total & Per unique \\
    \midrule
    GPT-6 Sol & Codex & 138 & 116 & 84 & 12 & 13 & 15 & 9 & 69 & 23 & 296{,}500 & 4{,}318 & 37 \\
Claude Opus 5 & Claude Code & 65 & 56 & 86 & 9 & 9 & 10 & 8 & 18 & 20 & 160{,}000 & 9{,}788 & 175 \\
Grok 4.7 & Grok Build & 42 & 25 & 60 & 6 & 6 & 4 & 1 & 13 & 7 & 57{,}500 & 4{,}120 & 165 \\
DeepSeek V4.1 Flash & Codex & 24 & 19 & 79 & 6 & 6 & 5 & 4 & 4 & 6 & 68{,}000 & 208 & 11 \\
Kimi K3 & Codex & 32 & 15 & 47 & 7 & 4 & 1 & 1 & 7 & 6 & 30{,}000 & 2{,}478 & 165 \\
GLM 5.3 & Codex & 26 & 13 & 50 & 5 & 3 & 3 & 2 & 2 & 6 & 38{,}000 & 5{,}020 & 386 \\
GPT-6 Luna & Codex & 7 & 4 & 57 & 4 & 1 & 0 & 1 & 1 & 2 & 11{,}000 & 92 & 23 \\
\midrule
All models & & 334 & 248 & 74 & 12 & 42 & 38 & 26 & 114 & 70 & 661{,}000 & 26{,}023 & 105 \\
    \bottomrule
  \end{tabular}
\end{table*}

\PP{Open-weight models trail flagship models substantially}
\sol finds 116 unique bugs, followed by \opus with 56 and \grok with 25.
The three open-weight models find 13--19 each, and \luna finds 4.
In its first 1M output tokens per campaign, \sol finds 51 unique bugs for \$1.5K. Of the other models, only \opus exceeds that count at 3M, finding five more bugs for 6.7$\times$ the cost (\autoref{f:discovery}).
\sol costs about as much as \grok (\$4.3K versus \$4.1K) and finds 4.6$\times$ as many unique bugs (\autoref{f:discovery}b).
\flash has the lowest discovery cost per unique bug (\$11). Its 19 bugs have a higher total estimated bounty than \grok's 25 (\$68{,}000 versus \$57{,}500) at a twentieth of the discovery cost.
RCE and EoP account for 47\% of \flash's findings and 20\% of \grok's.

Report precision is 79--86\% for \sol, \opus, and \flash and 47--60\% for the other four models.
Of the 31 reports from \sol and \opus that add no unique finding, 20 are valid duplicates. Of the 55 such reports from the other five models, 47 fail validation (\autoref{s:rq3}).

Across successive intervals of 1M output tokens per campaign, \grok finds 18, 6, and 1 unique bugs (\autoref{f:discovery}a). It submits 18, 12, and 12 reports in those intervals, so its precision falls from 100\% to 50\% to 8\%.
\flash finds 10, 7, and 2, with precision of 100\%, 88\%, and 33\%.
In the final 1M interval, \opus finds 14 more unique bugs, \kimi 8, \glm 5, and \luna none.
Interestingly, in that interval alone, \sol finds 43 unique bugs, more than \grok does over the full 3M output tokens per campaign.

\begin{figure*}[t]
  \centering
  \includegraphics{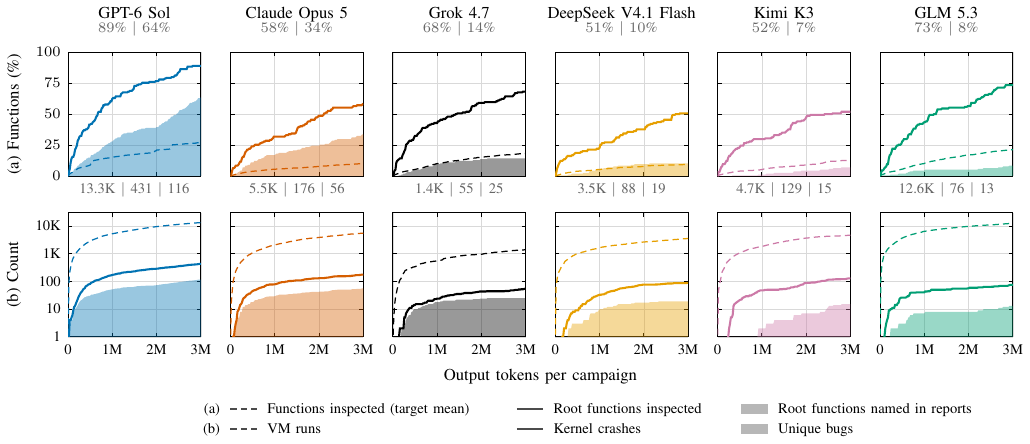}
  \caption{Function inspections and reported findings for six of the seven models (\luna omitted).
    In (a), the dashed line gives the fraction of target functions inspected, averaged over the 12 targets.
    The solid line and shaded area use the 154 root functions identified across all seven models as their denominator.
    They show the fractions with a recorded inspection and with a unique report from the displayed model, respectively.
    The headings in (a) list these two fractions at 3M output tokens per campaign.
  At the same budget, \sol's heading in (b), \mbox{13.3K $\mid$ 431 $\mid$ 116}, denotes 13{,}283 VM runs (dashed), 431 kernel crashes (solid), and 116 unique bugs (shaded) across its 12 campaigns.}
  \label{f:coverage}
\end{figure*}

\begin{figure*}[t]
  \centering
  \includegraphics{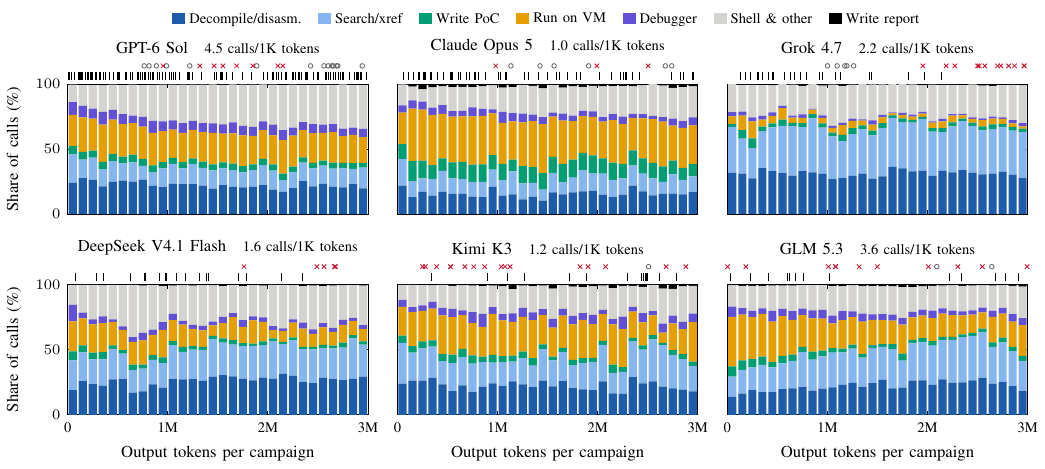}
  \caption{Tool calls by activity for six of the seven models (\luna omitted).
    Shares are calculated for each 100K output tokens and averaged over the model's 12 campaigns.
  Above each panel, ticks mark unique reports, circles valid duplicates, and crosses hard fails.}
  \label{f:activity}
\end{figure*}

\PP{Function inspection and VM runs do not strongly correlate with bug finding}
The 248 findings identify 154 root functions. Across all seven models, recorded inspections cover 47--89\% of these functions. For comparison, each model inspects only 9--27\% of a target's full function set on average.
Even among root functions that a given model never names in a unique report, its logs record inspections for 40--71\%, depending on the model.
This is 3.0--6.9$\times$ the model's inspection rate for all functions after weighting each target by its number of such root functions.
\glm has recorded inspections for 113 of the 154 functions and names 13 in unique reports. For \opus the counts are 90 and 52 (\autoref{f:coverage}a).
\luna's logs show inspections of 73 root functions, and its unique reports name 4.
Across the 84 campaigns, the number of unique bugs found has a Spearman rank correlation of $-0.04$ with the number of distinct target functions inspected and $+0.29$ with the number of VM runs.
Twelve campaigns have incomplete inspection records, so their function counts are lower bounds.

Let $R$, $C$, and $U$ denote VM runs, observed kernel crashes, and unique bugs for one model, summed over its 12 campaigns (\autoref{f:coverage}b).
With hit rate $h=100C/R$ and unique bugs per 100 crashes $y=100U/C$, the counts satisfy
\begin{equation}
  U = R \times \frac{h}{100} \times \frac{y}{100}.
  \label{e:discovery-decomposition}
\end{equation}
\glm and \sol make similar numbers of VM runs (12.6K and 13.3K) but record 76 and 431 crashes, respectively.
\grok makes only 1.4K VM runs and records 4.0 crashes per 100 runs, the highest hit rate. Static analysis accounts for 65\% of its tool calls and the target VM and debugger together for 7\% (\autoref{f:activity}).
\kimi records 129 crashes, compared with \grok's 55, but reports 12 unique bugs per 100 crashes, compared with 45 for \grok.
For findings with a recorded first report and a prior inspection of the root function, we count output tokens from first inspection to first report. The median is 931K for \kimi and ranges from 127K to 228K for \sol, \opus, \grok, and \flash.
Ten of \kimi's 16 hard fails exceed the replay build time limit (\autoref{s:rq3}).

\PP{Three parsers account for 75\% of findings}
The three disk-image parsers (i.e., \cc{ntfs.sys}, \cc{refs.sys}, and \cc{spaceport.sys}) account for 186 of the 248 unique bugs and 105 of the 154 root functions (\autoref{t:rq1-targets}).
Each model reports 50--85\% of its unique bugs in these parsers.
Among functions inspected by at least one model, unique reports name 33 root functions per 1{,}000 in the parsers and 3.2 per 1{,}000 elsewhere.
Out-of-bounds reads account for 54\% of the findings in the parsers and 16\% in the other targets.
In the parsers, \sol and \opus record 7.4 and 9.3 crashes per 100 VM runs, respectively, compared with 0.8 and 1.5 across the other targets.
\sol reports 99 unique bugs in the parsers and \opus 39. Each reports 17 in the other nine targets.
Across all targets, both models average 3.2 crashes per 100 VM runs. \sol runs 2.4$\times$ as many experiments, spending 2.7K output tokens per VM run against 6.5K for \opus.
Outside the parsers, \sol still runs twice as many experiments, but the models record similar crash counts (71 and 63).

\subsection{RQ2: Effects of Analysis Tool Support}
\label{s:rq2}

RQ2 runs \sol, RQ1's strongest model, under five tool conditions.
Three vary the disassembler (Ghidra as in RQ1, IDA Pro, or Binary Ninja), each with symbols.
The other two start from the Ghidra condition. \emph{No symbols} strips the targets' symbols. \emph{No disassembler server} removes Ghidra's decompiler, disassembly, cross-references, and search. It retains WinDbg, which can still disassemble, and the agent's shell. As discussed earlier, due to the high cost,
every campaign stops at 2M output tokens. Each disassembler has three runs of 12 campaigns, and each removal has one run.
Ghidra's first run reuses \sol's RQ1 campaigns, cut at 2M and judged anew with RQ2.

\begin{figure}[t]
  \centering
  \includegraphics{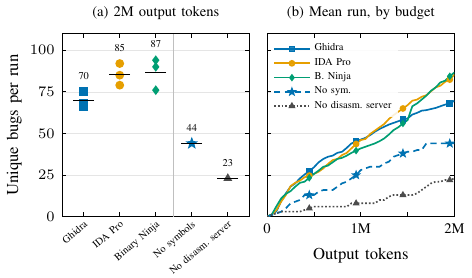}
  \caption{\sol reports more unique bugs with IDA Pro or Binary Ninja than with Ghidra, and fewer when symbols or the disassembler server are removed.
    (a)~Unique bugs per run of 12 campaigns at 2M output tokens per campaign, one dot per run and a bar at the mean.
  (b)~Mean unique bugs per run by output tokens spent per campaign.}
  \label{f:rq2-tools}
\end{figure}

\begin{table}[t]
\caption{\sol's activity per campaign at 2M output tokens.
  Function counts use disassembler-server results. First unique gives the median token count at the first unique report among campaigns with at least one.}
\label{t:rq2-mech}
\centering\footnotesize
\setlength{\tabcolsep}{2.5pt}
\begin{tabular}{@{}lrrrrrr@{}}
  \toprule
  Condition & \makecell{Functions\\read} & \makecell{Static\\calls (\%)} & \makecell{VM\\runs} & Crashes
    & \makecell{Hit\\rate} & \makecell{First\\unique} \\
  \midrule
  Ghidra & 812 & 37 & 774 & 21.2 & 2.7 & 160K \\
IDA Pro & 796 & 36 & 741 & 23.3 & 3.1 & 227K \\
Binary Ninja & 1079 & 45 & 696 & 20.0 & 2.9 & 276K \\
No symbols & 629 & 24 & 936 & 12.9 & 1.4 & 274K \\
No disasm.\ server & -- & 4 & 1011 & 10.1 & 1.0 & 651K \\
  \bottomrule
\end{tabular}
\end{table}

\PP{IDA Pro and Binary Ninja yield more bugs than Ghidra}
\sol averages 85 unique bugs per run with IDA Pro and 87 with Binary Ninja, compared with 70 with Ghidra (+22\% and +24\%, \autoref{f:rq2-tools}a).
The lowest counts with IDA Pro and Binary Ninja (79 and 76) both exceed the highest with Ghidra (75).
Each yields a higher mean than Ghidra on 8 of the 12 targets. Two targets, \cc{ntfs.sys} and \cc{win32k.sys}, account for about two thirds of each increase.
When we resample targets, the 95\% bootstrap intervals for the ratios to Ghidra both include 1 (0.97--1.93 and 0.98--2.00, respectively).

\PP{Fewer VM runs with IDA Pro and Binary Ninja}
At 1M output tokens, \sol averages 46 unique bugs with Ghidra, 46 with IDA Pro, and 41 with Binary Ninja (\autoref{f:rq2-tools}b).
The advantage over Ghidra appears in the second million tokens. By 2M, \sol has performed fewer experiments with IDA Pro and Binary Ninja than with Ghidra (741 and 696 VM runs per campaign against 774, \autoref{t:rq2-mech}).
The corresponding hit rates are 3.1, 2.9, and 2.7 crashes per 100 VM runs.
\sol reports 31 unique bugs per 100 crashes with IDA Pro, 36 with Binary Ninja, and 27 with Ghidra.
In \cc{ntfs.sys} and \cc{win32k.sys}, Ghidra's hit rates of 2.5 and 1.3 rise to 3.0 and 4.6 with IDA Pro and to 3.7 and 5.2 with Binary Ninja.

\PP{Ghidra names the fewest structure fields}
All three servers let \sol decompile, disassemble, follow references, read memory, and search.
For the 3{,}873 functions that all three decompile in full, Ghidra names the field in 2\% of fixed-offset memory accesses, compared with 14\% for IDA Pro and 9\% for Binary Ninja.
Ghidra writes nearly all of them as raw offsets, including the field IDA Pro calls \cc{CurrentStackLocation}.
Ghidra also puts a cast on 31\% of its lines and IDA Pro on 24\%, while Binary Ninja writes almost none.

\PP{Removing symbols reduces discoveries by 37\%}
Without symbols, \sol reports 44 unique bugs per run, 37\% below the Ghidra mean (95\% bootstrap interval for the ratio 0.46--0.96).
The decrease occurs in filesystem drivers. Outside them, the ratio is 1.12 (0.56--2.62).
In those drivers, \sol performs more VM runs per campaign without symbols (1{,}362 versus 938), but its hit rate falls from 4.0 to 1.5.
It reports 29 unique bugs per 100 crashes in these drivers without symbols and 27 with symbols. Overall report precision without symbols is 92\%, with no hard fails.

\PP{Removing the server reduces discoveries by 67\%}
Without the disassembler server, \sol reports 23 unique bugs per run, 67\% below the Ghidra mean (95\% bootstrap interval for the ratio 0.17--0.43).
All 23 reports are unique. Six of the 12 campaigns find no bugs. Among campaigns with a unique report, the median first report occurs at 651K output tokens, compared with 160K with Ghidra.
\sol performs 1{,}011 VM runs per campaign, compared with 774 with Ghidra. The hit rate falls from 2.7 to 1.0, and unique bugs per 100 crashes fall from 27 to 19.

\subsection{RQ3: Judging Reliability and Scoring}
\label{s:rq3}

\begin{figure}[t]
  \centering
  \includegraphics{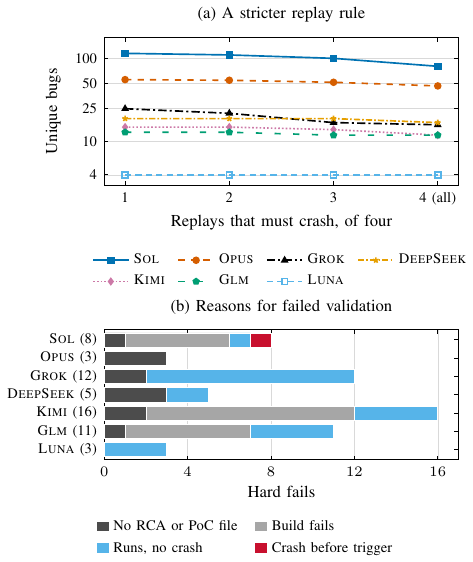}
  \caption{Replay outcomes for all seven RQ1 models.
    (a)~Unique findings retained when at least one associated PoC crashes the kernel in at least $k$ of four replays, on a log scale. The Judge uses $k=1$.
  (b)~Reasons for rejected reports.}
  \label{f:rq3-replays}
\end{figure}

We first examine how often the Judge reproduces a reported crash across four replays, using all 334 reports from RQ1 (\autoref{f:rq3-replays}). We then use reports with known answers to test whether the Judge accepts valid reports, rejects incorrect analyses, recognizes duplicates, and assigns the correct impact class (\autoref{t:rq3-scorecard}).

We compare the single-shot Judge used in RQ1 (GPT-6 Sol) with single-shot Claude Opus 5 and with an agentic GPT-6 Sol Judge that has access to Ghidra.
Thirty public reports of kernel vulnerabilities from 2020 to 2026 provide both valid cases and vendor impact labels. None of these vulnerabilities was found in RQ1. For each report, we supply the root-cause analysis and callstack and withhold the PoC.
To test rejection of invalid reports, we pair 30 real RQ1 crashes with deliberately incorrect analyses. To test deduplication, we use 48 reports describing ten defects patched in 2026.

A \emph{query} asks one Judge to assess one report. For validity and impact, we group three queries on the same evidence into one \emph{round} and take the majority verdict. This lets two agreeing judgments determine the verdict even when the third differs. For deduplication, one query supplies the round verdict.
Each Judge completes three rounds per report to test whether its verdict changes. Accuracy is the fraction of individual queries that agree with ground truth. Consistency is the fraction of reports that receive the same verdict in all three rounds.

\PP{Some valid PoCs crash intermittently}
Of the 334 reports, 276 were accepted after review. These comprise 248 unique findings and 28 valid duplicates.
For 186 of the 248 unique findings, the representative report's PoC crashes the kernel in all four replays. For the other 62, it crashes in one to three.
Across all accepted reports, including valid duplicates, choosing one replay per report at random would reproduce a crash for about 89\% of reports on average.
A stricter rule requires a single PoC to crash the kernel in all four replays. Counting either the representative PoC or one from a merged duplicate retains 189 findings and preserves the model ranking by estimated bounty (\autoref{f:rq3-replays}a).
Of the 276 accepted reports, 14 concern \cc{ntoskrnl.exe}, which we test only with Driver Verifier disabled (\autoref{s:design-judge}). We test the remaining 262 under both settings. Five reproduce the crash only with Verifier enabled and 19 only with it disabled.

Rejection reasons come from the first replay (\autoref{f:rq3-replays}b). Of the 58 rejected reports, 12 lack a required file, 21 have build failures, and 24 have no verified crash. In the remaining case, the crash precedes the operation cited as its cause.
Across all replays, 56 reports never crash. The two that crash are the preceding case and one report missing its root-cause analysis.
Ten of \kimi's 16 rejections are builds that exceed the replay time limit. All three of \opus's rejections lack a required file.

\providecommand{\score}[3]{\cellcolor{okorange!#1}#2\%\,{\tiny\color{black!60}#3}}
\providecommand{\noscore}{\textcolor{gray}{--}}
\begin{table}[t]
\caption{Accuracy is the share of queries that match ground truth. Consistency is the share of reports with the same verdict in all three rounds. The same 30 public reports supply the valid cases and the vendor impact labels.}
\label{t:rq3-scorecard}
\centering\footnotesize
\setlength{\tabcolsep}{2.7pt}
\begin{tabular}{@{}lcccc}
  \toprule
  & \multicolumn{2}{c}{Validity} & Uniqueness & Impact \\
  \midrule
  \rowcolor{black!8}\multicolumn{5}{c}{\emph{Test set}} \\
\quad Reports & 30 & 30 & 48 & 30 \\
\quad Bugs & 30 public & 30 in RQ1 & 10 patched & 30 public \\
\quad Ground truth & valid & invalid & same bug & vendor's \\
\quad Rounds per report & 3 & 3 & 3 & 3 \\
\quad Queries per round & 3 & 3 & 1 & 3 \\
\rowcolor{black!8}\multicolumn{5}{c}{\emph{Accuracy: queries that match the ground truth}} \\
\multicolumn{5}{@{}l}{\quad Single-shot Judge} \\
\qquad GPT-6 Sol & \score{0}{100}{270/270} & \score{0}{100}{270/270} & \score{0}{100}{144/144} & \score{43}{95.2}{257/270} \\
\qquad Claude Opus 5 & \score{0}{100}{270/270} & \score{13}{98.5}{266/270} & \score{0}{100}{144/144} & \score{47}{94.8}{256/270} \\
\multicolumn{5}{@{}l}{\quad Agentic Judge} \\
\qquad GPT-6 Sol & \score{0}{100}{270/270} & \score{17}{98.1}{265/270} & \score{0}{100}{144/144} & \score{60}{92.6}{250/270} \\
\rowcolor{black!8}\multicolumn{5}{c}{\emph{Consistency: reports with one round verdict in all three rounds}} \\
\multicolumn{5}{@{}l}{\quad Single-shot Judge} \\
\qquad GPT-6 Sol & \score{0}{100}{30/30} & \score{0}{100}{30/30} & \score{0}{100}{48/48} & \score{0}{100}{30/30} \\
\qquad Claude Opus 5 & \score{0}{100}{30/30} & \score{30}{96.7}{29/30} & \score{0}{100}{48/48} & \score{0}{100}{30/30} \\
\multicolumn{5}{@{}l}{\quad Agentic Judge} \\
\qquad GPT-6 Sol & \score{0}{100}{30/30} & \score{0}{100}{30/30} & \score{0}{100}{48/48} & \score{60}{93.3}{28/30} \\
  \bottomrule
\end{tabular}
\end{table}

\PP{Single-shot Sol rejects all 30 invalid reports}
Each configuration accepts the valid public reports in all 270 queries (30 reports $\times$ 3 rounds per report $\times$ 3 queries per round).
The separate set of 30 invalid reports likewise yields 270 queries per configuration. RQ1's Judge had accepted the original analysis for each crash. Of the 30 altered analyses, ten blame a different defect in the same binary, ten blame a defect in another binary, and ten place the fault at a step inconsistent with the crash evidence.
Single-shot GPT-6 Sol rejects the altered analysis in all 270 queries. Claude Opus 5 accepts the same incorrect analysis in four queries, producing one incorrect majority verdict in 90 rounds. The agentic Judge accepts an incorrect analysis in five queries, each time citing a real defect different from the one claimed. Its majority verdict rejects the analysis in all 90 rounds.

\begin{figure}[t]
  \centering
  \includegraphics{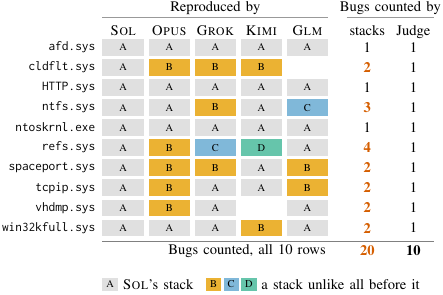}
  \caption{Reproductions of ten patched defects, one defect per row. Within a row, matching normalized callstacks share a letter. Normalization removes offsets, user-mode frames, and frames for crash reporting or pool allocation. A truncated stack can match the start of a longer one.}
  \label{f:rq3-dedup}
\end{figure}

\PP{All three Judges identify the tested duplicates}
Seven of the ten defects produce multiple normalized callstacks.
For 15 of the other models' 38 reproductions, the stack differs from \sol's for the same defect. Some enter through another system call and others crash at another point on the same path.
Grouping reproductions by callstack within each defect gives 20 groups in total (\autoref{f:rq3-dedup}).
Each Judge identifies all 48 reproductions as duplicates in all three rounds (144 correct judgments per Judge, 432 in total).
All three Judges recognize duplicates even when their callstacks differ.

\PP{Single-shot Judges match more vendor labels}
The impact study's 30 reports include 19 EoP and four RCE vulnerabilities.
Single-shot GPT-6 Sol matches the vendor's impact label in 257 of 270 queries (95.2\%) and Claude Opus 5 in 256 (94.8\%).
The agentic Judge matches in 250 (92.6\%).
Each single-shot Judge assigns the same impact class in all three rounds for every report. In one case, GPT-6 Sol consistently assigns the wrong class. The agentic Judge changes its verdict on two of the 30 reports.
All 47 incorrect query responses classify an EoP vulnerability as ID. In its additional errors, the agentic Judge cites evidence of a read and finds insufficient evidence of a write or control-flow effect.
It uses about 49$\times$ as many input tokens per query as single-shot GPT-6 Sol on average (870.6K versus 17.9K).

\subsection{RQ4: Zero-Days and Their Impact}
\label{s:rq4}

We conducted the main evaluation in September 2026 and use that month's cumulative update, Windows build 26100.9457, as the reference for 0-days.
For each unique finding, we replay its PoC on this common reference build, nine months newer than the discovery build.
Each PoC runs as a standard user with and without Driver Verifier, or twice without it for \cc{ntoskrnl.exe}.
We classify a finding as a \emph{0-day} if the PoC still crashes the reference build and the kernel debugger captures a symbolized trace.

\providecommand{\notapp}{\textcolor{black!40}{--}}
\begin{table}[t]
\caption{Over two in five unique bugs still crash a Windows build nine months newer.
Rows labeled ``Across'' count each 0-day defect once across the indicated models or runs.}
\label{t:rq4-zeroday}
\centering\footnotesize
\setlength{\tabcolsep}{3pt}
\begin{tabular}{@{}c@{\hspace{4pt}}lrrrrrrr@{}}
\toprule
 & \multirow{2}{*}[-1pt]{Source} & \multirow{2}{*}[-1pt]{Unique} & \multicolumn{2}{c}{0-days} & \multicolumn{4}{c}{By impact} \\
\cmidrule(lr){4-5}\cmidrule(l){6-9}
 & & & Count & \% & RCE & EoP & ID & DoS \\
\midrule
\multirow{9}{*}{\rotatebox[origin=c]{90}{RQ1}} & GPT-6 Sol & 116 & 46 & 40 & 3 & 1 & 26 & 16 \\
 & Claude Opus 5 & 56 & 23 & 41 & 1 & 1 & 8 & 13 \\
 & Grok 4.7 & 25 & 11 & 44 & 0 & 0 & 5 & 6 \\
 & DeepSeek V4.1 Flash & 19 & 13 & 68 & 2 & 2 & 3 & 6 \\
 & Kimi K3 & 15 & 11 & 73 & 0 & 0 & 7 & 4 \\
 & GLM 5.3 & 13 & 5 & 38 & 0 & 0 & 1 & 4 \\
 & GPT-6 Luna & 4 & 1 & 25 & 0 & 0 & 0 & 1 \\
\cmidrule(l){2-9}
 & Sum over models & 248 & 110 & 44 & 6 & 4 & 50 & 50 \\
 & Across models & \notapp & 70 & \notapp & 5 & 4 & 29 & 32 \\
\midrule
\multirow{7}{*}{\rotatebox[origin=c]{90}{RQ2}} & Ghidra $\times$3 & 209 & 83 & 40 & 5 & 3 & 33 & 42 \\
 & IDA Pro $\times$3 & 256 & 107 & 42 & 5 & 4 & 32 & 66 \\
 & Binary Ninja $\times$3 & 260 & 100 & 38 & 8 & 4 & 28 & 60 \\
 & No symbols & 44 & 20 & 45 & 3 & 0 & 6 & 11 \\
 & No disasm.\ server & 23 & 14 & 61 & 0 & 0 & 6 & 8 \\
\cmidrule(l){2-9}
 & Sum over runs & 792 & 324 & 41 & 21 & 11 & 105 & 187 \\
 & Across runs & \notapp & 125 & \notapp & 11 & 7 & 37 & 70 \\
\midrule
 & Across RQ1 and RQ2 & \notapp & 146 & \notapp & 12 & 9 & 43 & 82 \\
\bottomrule
\end{tabular}
\end{table}

\PP{The two experiments identify 146 distinct 0-days}
In RQ1, PoCs for 110 of 248 within-model findings (44\%) still crash the September build (\autoref{t:rq4-zeroday}). In RQ2, the corresponding count is 324 of 792 within-run findings (41\%).
The 110 RQ1 findings span 11 of the 12 targets.
After deduplication across models, RQ1's 110 findings represent 70 unique defects.
Only 16 of these defects are found by more than one model.
Deduplication across runs reduces RQ2's 324 0-day findings to 125 unique defects.
We count the shared Ghidra baseline once. Deduplication across both experiments yields 146 distinct defects.
Of these, 76 appear only in RQ2, and 58 of those appear in a single run.

For 77 of RQ1's 110 findings, another model inspected the same root function on the same target by the output-token count at which the finding was reported.
That other model never submitted a unique report naming the function during its campaign.
For one of \glm's 0-days, all six other models inspected the root function within their first 58K output tokens. \glm reported the defect after about 600K output tokens.

\PP{Crash reproduction varies by impact class}
On the September build, PoCs still reproduce the reported crash for 6 of 38 RCE findings and 4 of 26 EoP findings, compared with 50 of 114 ID and 50 of 70 DoS findings.
For 48 of the 54 RCE and EoP findings whose PoCs no longer crash, the update repairs the reported defect itself. For each of the other six findings, the update blocks the PoC elsewhere on its path.
Of the 20 DoS findings whose PoCs no longer crash, 10 match a fix to the reported defect.

\PP{The main evaluation (RQ1) found only 58 of 228 1-day vulnerabilities}
Here we count a reference vulnerability as found if its fix blocks at least one RQ1 PoC. This criterion includes fixes that block the PoC elsewhere on its execution path.
The 228 \mbox{1-days} are vulnerabilities present in the evaluated December 2025 build and fixed by updates released from January through September 2026 (\autoref{t:targets}).
They provide a partial answer key.
Of the 58 matching fixes, 37 block a PoC from \sol.
The remaining 170 reference entries have no matching RQ1 finding.
A fix that blocks PoCs from several models counts once.

\begin{figure}[t]
  \centering
  \includegraphics{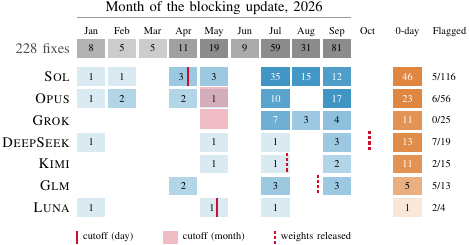}
  \caption{Monthly counts of \windiff fixes (gray) and RQ1 findings by the update that blocks their PoC (blue).
    The 0-day column counts findings reproducible on the September build. The October column marks only \flash's weights release.
  Flagged gives the number of potentially exposed findings out of each model's unique findings, under the date criteria in the text.}
  \label{f:rq4-patches}
\end{figure}

\PP{Update timing flags 27 of 248 findings}
A model that saw a bug's fix during training could rediscover the bug from memory.
For all 138 RQ1 findings whose PoCs no longer crash the September build, we matched the update that blocks the PoC by hand to a \windiff entry (\autoref{s:design-windiff}).
We flag findings if the update was public by the reporting model's knowledge cutoff, including the cutoff month when only the month is known.
For the three models without a published cutoff, we use the weights' release date as an upper bound (\autoref{f:rq4-patches}).
For example, five of \sol's 116 findings match an update released by its April 20 cutoff. The remaining 111 either match a later update or still reproduce on the September build.

This rule flags 26 findings with matched updates. In the later \flash evaluation, one PoC that crashed the September build no longer crashed the October update preceding the model's release.
We include that finding conservatively without attributing it to a specific fix.
Together these 27 findings (11\%) bound potential exposure to public updates. Their dates alone cannot establish whether a model learned from a fix.
Excluding all 27 leaves 221 findings. Only \kimi and \flash change rank, with 13 and 12 remaining findings, respectively.

For the four models with published cutoffs, we also compare fixes issued before and after each model's cutoff~\cite{jain2025livecodebench,roberts2024cutoff}.
Of the 228 fixes, 171 shipped from July to September, after every published cutoff.
For each \windiff entry, we ask whether its fix blocks at least one of that model's PoCs, omitting updates that cannot be ordered relative to the cutoff.
The model-stratified odds ratio for a match after versus before the cutoff is 0.92 (95\% CI 0.46--1.83).
This interval is consistent with both lower and higher odds of a match after the cutoff.

\section{Discussion and Limitations}
\label{s:discussion}

\PP{One kernel build}
\sys measures twelve kernel targets of one Windows build, so its counts describe Windows alone.
Its targets still cover files, system calls, and network traffic, the three ways untrusted input reaches the kernel (\autoref{t:targets}).
Porting \sys to another kernel needs its own VM image, debugger bridge, and impact table.

\PP{One campaign per model and target}
Cost keeps us from repeating every experiment three times, so RQ1 runs one campaign per model and target.
Across the seven models, unique-bug counts range from 4 to 116, a 29-fold difference. The five models using Codex span the same range.
In RQ2 we can estimate run-to-run variation for \sol. With each disassembler, the difference between its highest and lowest counts over three runs is at most 21\% of their mean.
The corresponding variation for the other models is unknown.

\PP{Cost of a run}
Across all 84 RQ1 campaigns, the first 1M output tokens per campaign cost \$8.4K, compared with \$26.0K for the full 3M.
The 1M budget already ranks \sol, \opus, \grok, and \flash in the same order as 3M.
The additional budget matters for measuring sustained discovery. In the final 1M, \sol finds 43 unique bugs, \grok one, and \flash two (\autoref{s:rq1}).

\PP{A model grades validity, uniqueness, and impact}
The Judge uses GPT-6 Sol, which also reports the most unique bugs. In the three ground-truth studies, Claude Opus 5 agrees with Sol on all round verdicts except one false acceptance.
We built the invalid reports ourselves, so their errors may differ from those agents make. The duplicate study does not test distinct defects with similar crashes.
On the discovery reports, human review overturns one of 277 acceptances in RQ1. It reverses none of 27 merges in RQ1 and seven of 70 in RQ2.
The review adds one merge in RQ1 and seven in RQ2 and corrects impact labels in both directions. We report the corrected counts.

\PP{Validation of 0-day findings}
Each 0-day in our count has a PoC that reproduces the crash on the September build with a kernel debugger trace. The Judge also checks the reported root cause against the replay evidence.

\PP{A prompt without strategy}
The prompt asks for every valid, unique crash (\autoref{f:prompt}) and leaves the search strategy to the agent. Directing agents to prioritize RCE or EoP could change both the number and the severity of the bugs they report.
We left hints out to give each model full autonomy, which lets RQ1 trace how models search a binary on their own. The resulting counts provide a baseline for experiments that specify a strategy.

\PP{Severity of the 0-days}
Of 64 RQ1 findings classified as RCE or EoP, ten still reproduce on the September build. For 48 of the other 54, the corresponding update repairs the reported defect itself.
Most of RQ1's 110 model-level 0-day findings are ID or DoS.
The estimated bounty weights findings by impact. An EoP finding contributes sixteen times as much as a DoS finding.

\section{Related Work}
\label{s:relwk}

\PP{LLM-assisted binary analysis}
One line of work recovers readable code from binaries. LLM4Decompile~\cite{llm4decompile} decompiles assembly. DeGPT renames, simplifies, and comments decompiled code~\cite{degpt}.
DecLLM~\cite{decllm} repairs decompiler output for recompilation and evaluates vulnerability detection with CodeQL.
Other work recovers names and types. ReSym~\cite{resym} recovers variable and structure names and types, and GenNm~\cite{gennm} predicts variable names using caller/callee context.
\sys measures how the choice of disassembler and the availability of symbols affect the bugs agents report.

\PP{Bug discovery in binaries}
Kernel fuzzers generate inputs suited to the interface under test. For example, NtFuzz~\cite{ntfuzz} infers Windows syscall types to generate calls. For filesystems, Janus~\cite{janus} mutates both disk-image metadata and file-operation sequences.
For coverage feedback, kAFL~\cite{kafl} uses Intel Processor Trace to record execution in filesystem drivers on Windows, Linux, and macOS.
Hydra~\cite{hydra} checks recovered images against the filesystem's permitted post-crash states and tests file operations for POSIX violations.

A second question is how to choose which code to investigate. \textsc{Slyp}~\cite{lee2026slyp} analyzes Windows COM services and kernel drivers from supplied entry points and a target vulnerability class.
FirmAgent~\cite{firmagent} uses fuzzing-derived taint sources to guide LLM analysis of IoT firmware. Veritas~\cite{veritas} checks out-of-bounds claims along source-to-sink flows.
Other systems provide tools to inspect code or test inputs. ClearAgent~\cite{clearagent} queries a binary language server.
\mbox{\textsc{ReDoSpector}}~\cite{redospector} fuzzes Windows services for resource leaks with LLM-assisted~seeds.
Bajaj et al.~\cite{bringagents} study one workflow across Windows drivers and web applications. In contrast, \sys holds the task fixed to compare models and tools. Every agent receives a target binary and build, and \sys supplies no entry points, vulnerability class, or candidate flows.
Our main model comparison also allows 3M output tokens per campaign for long-horizon exploration.

\PP{Security benchmarks for agents}
SEC-bench~\cite{secbench} and CyberGym~\cite{cybergym} evaluate vulnerability reproduction from reports. ExploitGym~\cite{exploitgym}, ExploitBench~\cite{exploitbench}, and KEX-bench~\cite{kexbench} grade exploitation of known bugs.
For discovery, SEC-bench Pro~\cite{secbenchpro} specifies source paths, vulnerability classes, and expected error types. MobileCybench~\cite{mobilecybench} checks Android exploits with security-property probes.
CyberGym-E2E~\cite{cybergyme2e} and FuzzingBrain-Bench~\cite{fuzzingbrainbench} evaluate open-source discovery without prescribing a target vulnerability.
Pre-cutoff tasks may be memorized~\cite{jain2025livecodebench,roberts2024cutoff}, so SRE-Bench~\cite{srebench} uses private programs written from scratch.
\sys dates updates to commercial kernel binaries by patch diffing~\cite{apeg,patchdiffai} to assess which updates could have been public before training.

\PP{Validating and deduplicating bugs}
SEC-bench Pro grades PoCs with an LLM judge, and FuzzingBrain-Bench counts distinct crash signatures.
A signature can split reports of one bug or group different defects together~\cite{klees2018,igor}.
Other methods group crashes by approximate fixes inferred from patch templates~\cite{semanticbucketing}, by similarity of minimized traces~\cite{igor}, or by LLM embeddings of reports~\cite{gptrace}.
Our LLM Judge checks each report's root-cause analysis against crash evidence from PoC replays to assess validity and impact.
It merges reports only when one minimal fix would block both triggers. The Judge infers what that fix would repair without applying it.

\section{Conclusion}
\label{s:concl}
\sys gives agents an unmodified commercial kernel, no hint of where bugs lie, and most of a day per campaign, and keeps only bugs that replay.
Agents still uncover 146 kernel defects that survive nine more months of updates.
Every model reads about half or more of the functions that hold bugs, but the models differ in how they experiment on them and how long they keep finding bugs.
Every closed-weight flagship finds more bugs than any open-weight model, though the cheapest open-weight model scores a higher estimated bounty than one at a twentieth of its cost.
\sys thus offers a way to measure long-running, unguided vulnerability hunting in complex binaries and to compare how models carry it out, not just what they find.

\section*{Acknowledgments}
OpenAI's Daybreak for Frontline Defenders grant supported our experiments.
We are especially grateful to Jiawei Liu from OpenAI for his support.
We also thank Vector~35 for providing Binary Ninja licenses.

\bibliographystyle{IEEEtran}
\bibliography{p}

\end{document}